\documentclass[12pt]{iopart}
\usepackage{iopams}
\usepackage{times}
\usepackage{graphicx}%

\begin{document}

\title[Superfluid atomic gyroscopes]{Gyroscopic
motion of superfluid trapped atomic condensates}

\author{Halvor M{\o}ll Nilsen \dag, \ Dermot McPeake \ddag $^*$ and J F McCann  \ddag}

\address{\dag\
Institute of Physics, University of Bergen, All\'egaten 55,
N-5007
Bergen, Norway}

\address{\ddag\
Dept of Applied Mathematics and Theoretical Physics,
Queen's University Belfast,
Belfast, BT7 1NN,
Northern Ireland}

\address{
$^{*}$ NMRC, University College Cork,
Lee Maltings,
Prospect Row,
Cork,
Ireland.}

%\ead{j.f.mccann@qub.ac.uk}

\begin{abstract}

The gyroscopic motion of a trapped Bose gas containing a vortex is studied.  
 We model the system as a classical top, as a 
 superposition of coherent hydrodynamic
states, by solution of the Bogoliubov equations, 
and  by integration of  the
time-dependent Gross-Pitaevskii equation.
The frequency spectrum  of 
Bogoliubov excitations, including 
quantum frequency shifts,  
is calculated  and the quantal precession 
frequency is  found to be 
consistent with experimental results, though a small 
discrepancy exists. The superfluid precession 
is found to be well described by the classical
and hydrodynamic models.  However
the  frequency shifts and 
helical oscillations associated with vortex bending 
and twisting require
a quantal treatment. 
In gyroscopic precession, the vortex excitation modes 
$m=\pm 1$ are the dominant 
features giving a vortex kink or bend, while the
$m=+2$ is found to be the dominant Kelvin wave 
associated with vortex twisting.

\end{abstract}

\maketitle

\section{Introduction}

Superfluidity is one of the most dramatic illustrations of the quantum
nature of matter. It is exemplified in its purest form
 in  Bose-Einstein
condensation  \cite{bec} of cold atomic gases. The
quantisation of angular momentum associated with the phase
of the quantum fluid has been established
 through the  nucleation
of  vortices \cite{vort1,vort2} recently  extending to
the observation of 
large vortex lattice structures with grain boundaries \cite{vort3}.
Experiments have also revealed collective irrotational 
oscillatory flows  in the superfluid phase
\cite{quench1,quench2} that revert to thermal flow 
modes and frequencies  above the critical temperature.

\begin{figure}
{\centering
\includegraphics[width=10cm]{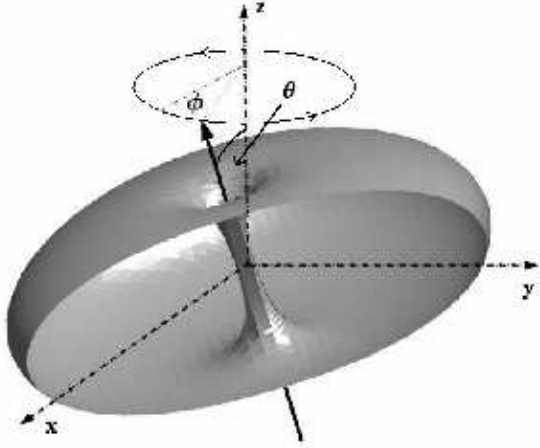}}
\caption{
  Classical rigid gyroscopic motion of the inhomogeneous 
   trapped condensate. The
figure  indicates a 
surface of constant density for a singly-quantised
gyroscope ($\kappa=1$) corresponding to the ground state 
solution of the Gross-Pitaevskii equation 
with atom interaction strength $C=1000$. 
The classical Euler angles \cite{goldstein}, $\theta$ and $\phi$, 
defining the orientation of the vortex are shown.}
\label{figure_1}
\end{figure}

One of the fundamental
properties of a pure quantum fluid is the coherence of the phase
of the wavefunction \cite{pines}  where different regions
of the condensate are phase-locked in their motion. For 
example, in the flow of an 
condensate past an  obstacle,  the requirement
of upstream and downstream continuity in mass transport and phase gradient
explains the capacity for superfluid flow \cite{flow_review},
and the transition to dissipation when local phase slip arises.
 For homogeneous and inhomogeneous
condensates, there is conclusive evidence, both theoretical and
experimental, that the nucleation of vortices is the
primary  mechanism responsible for
the onset of drag and dissipation in condensate flow
\cite{flow_review,flow,critical}.
Phase coherence  is responsible for
the quantisation of circulation (angular momentum) and the capacity
of the fluid to conserve its state of rotation unless acted upon by a critical
external moment. Consequently, it was  predicted \cite{vort_exfq,string_gyro} 
that a Bose-condensed gas could be made
to exhibit gyroscopic motion. Very recently this phenomenon 
has been observed experimentally \cite{ho02} and measured 
for the first time. 
In analogy with a spinning top, slightly displaced from
equilibrium by an impulsive moment,  a slight tilt of the
axis of rotation (figure \ref{figure_1})  produces precessional
and nutational motion.
In this paper we show that internal flows past the vortex 
create a Magnus force which results in precession, and that this combined 
with the inhomogeneity  in the 
condensate gives rise to vortex bending and twisting  into 
helical modes of oscillation.

The quantum gyroscope is also a  unique diagnostic tool
able to probe condensate flow velocities, acoustic waves, and pressures. 
The oscillating and twisting  motion of a vortex, as we
will show, is highly sensitive to density fluctuations and
inhomogeneities. In this paper we study the dynamics of a
single-quantum vortex lying near the axis of a
cylindrically symmetric trap. In our study we have employed four different
methods of analysing the motion:
(a) the Bogoliubov-de Gennes equations
(b) the Gross-Pitaevskii equation, (c) hydrodynamic equations, and  (d)
classical equations. While methods (a) and (c) are 
essentially linearisations of (b), they 
are extremely important in identifying the underlying modes 
of excitation.  
We study the collective
modes of the system and compare with recent hydrodynamic
theory. We find that while many features of the gyroscopic motion
are  described by the hydrodynamic model, the vortex motion
is significantly different. Firstly, the frequencies
of precessional motion are shifted from the hydrodynamic predictions, 
secondly the vortex core has a single kink, and thirdly the vortex
performs a helical Kelvin wave oscillation
of mode $m=+2$ not previously seen.

\section{Formulation of the problem}\label{sec:theory}

For a cold weakly-interacting gas, the ground state 
(condensate mode) dominates the collective
dynamics of the system. In experimental realizations one can
achieve temperatures such that $T \ll T_c$ (typically $0.1$ to
$1\mu{\rm K}$) and densities such that the gas is weakly 
interacting and highly dilute.  Under such conditions, the
condensate of $N_0 \gg 1$ atoms is well described by 
a mean field, or  wavefunction, 
  governed by the Gross-Pitaevskii equation, and
the quasiparticle excitations are acoustic
waves within this field. If the perturbations of the condensate 
are small, then it is appropriate and convenient to use
the linear response approximation, which is equivalent to
the Bogoliubov  approximation for single-particle
excitations in highly-condensed quantised Bose gases at zero
temperature.

\begin{figure}
{\centering
\includegraphics[width=5cm,angle=0]{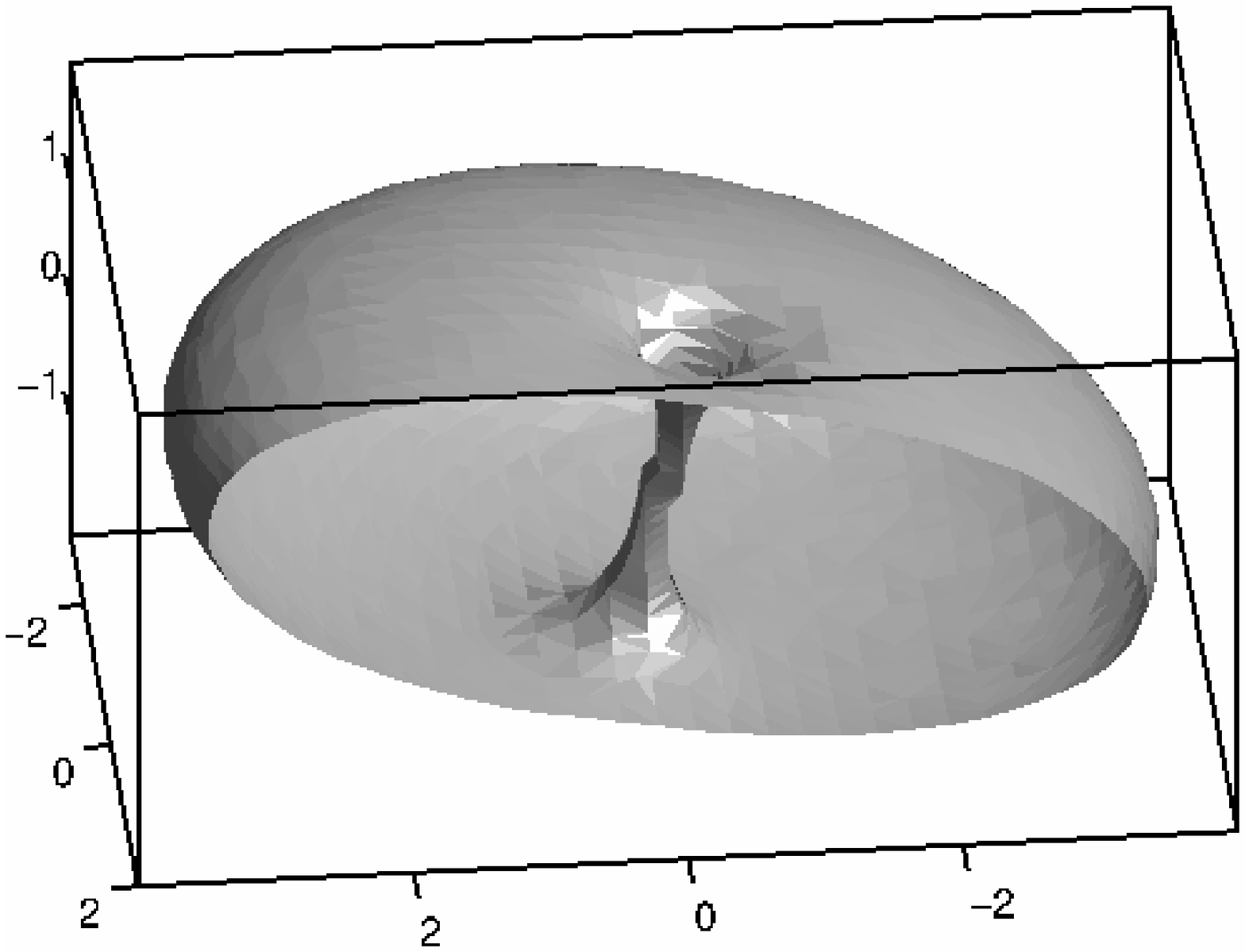}
\includegraphics[width=5cm,angle=0]{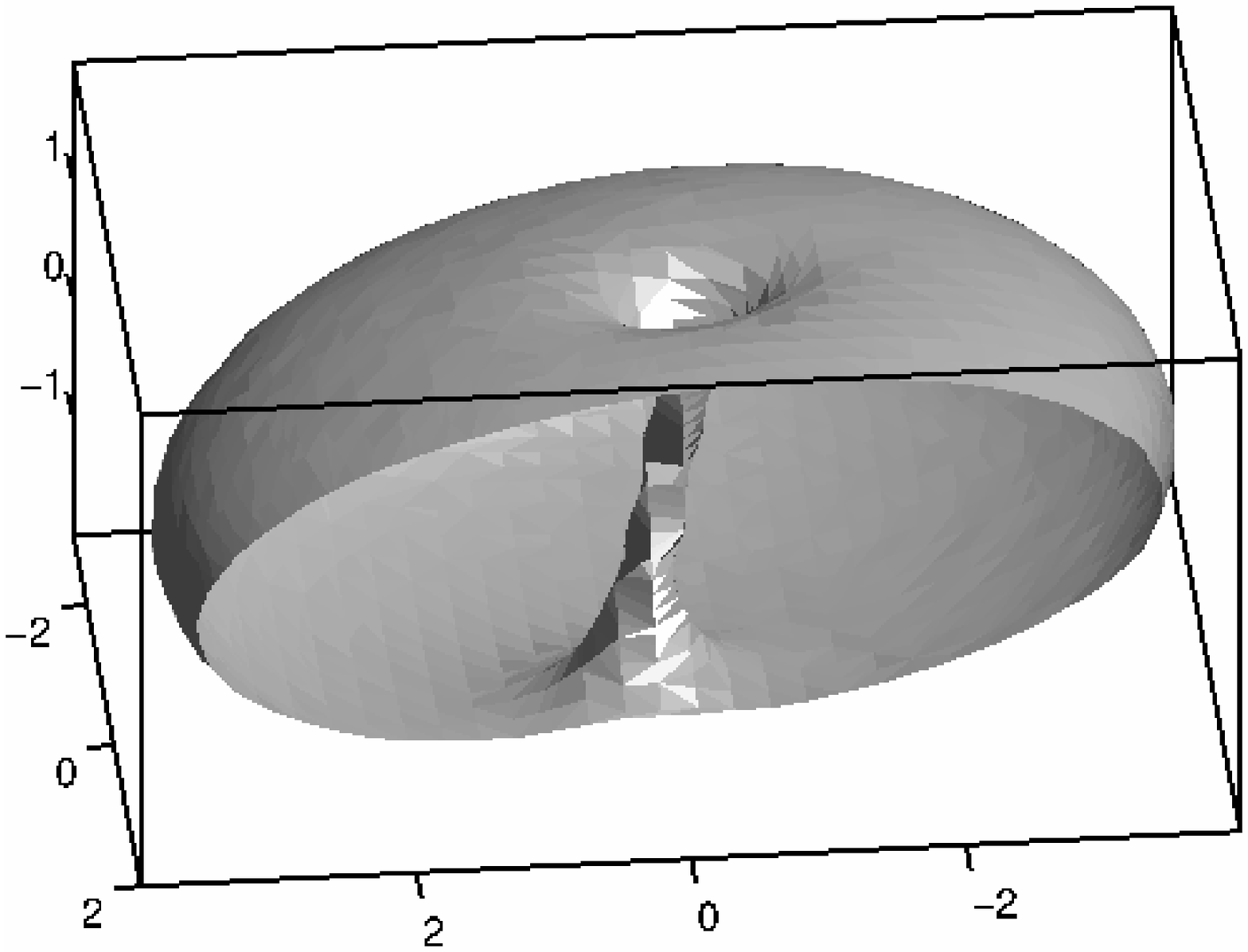}
\includegraphics[width=5cm,angle=0]{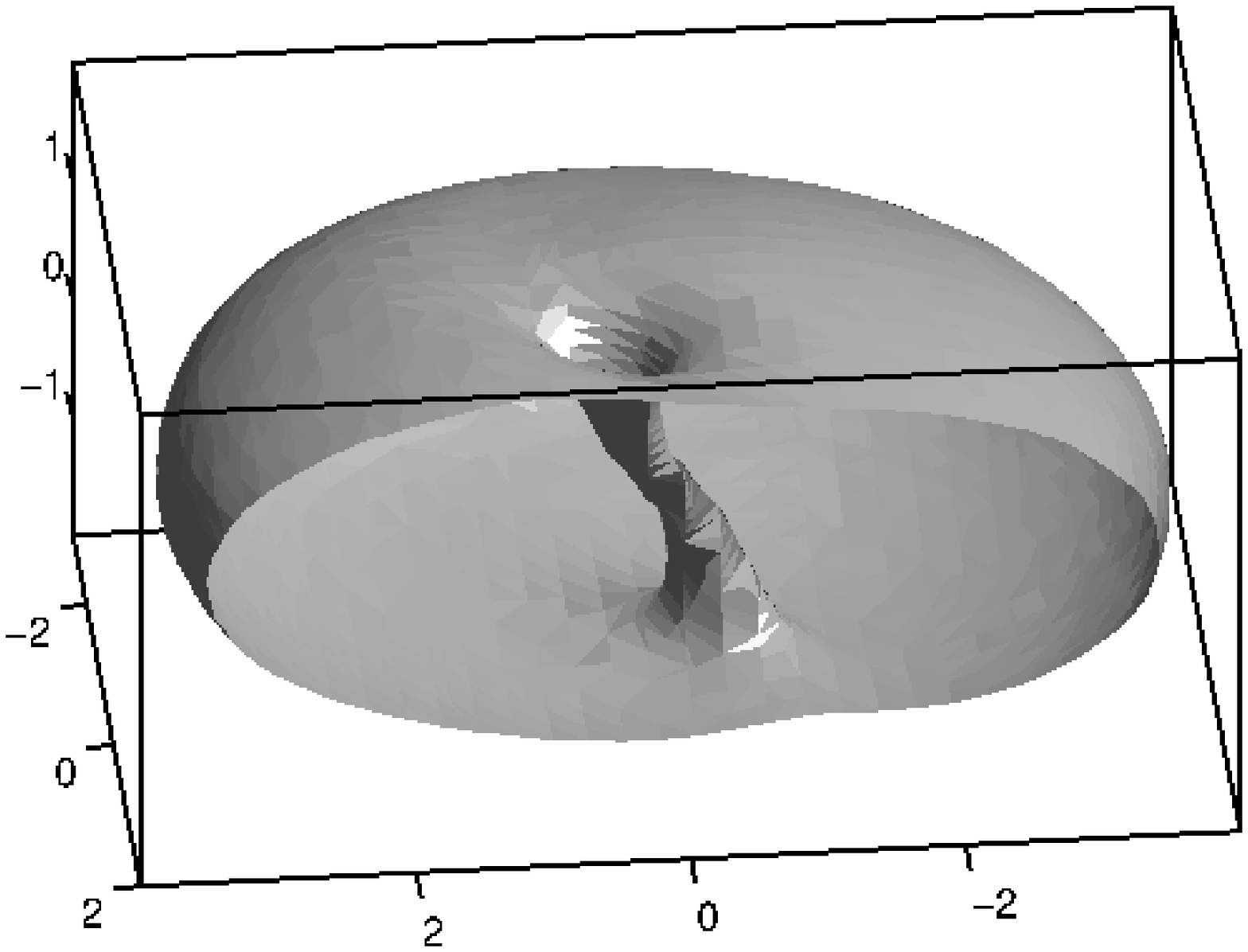}
\includegraphics[width=5cm,angle=0]{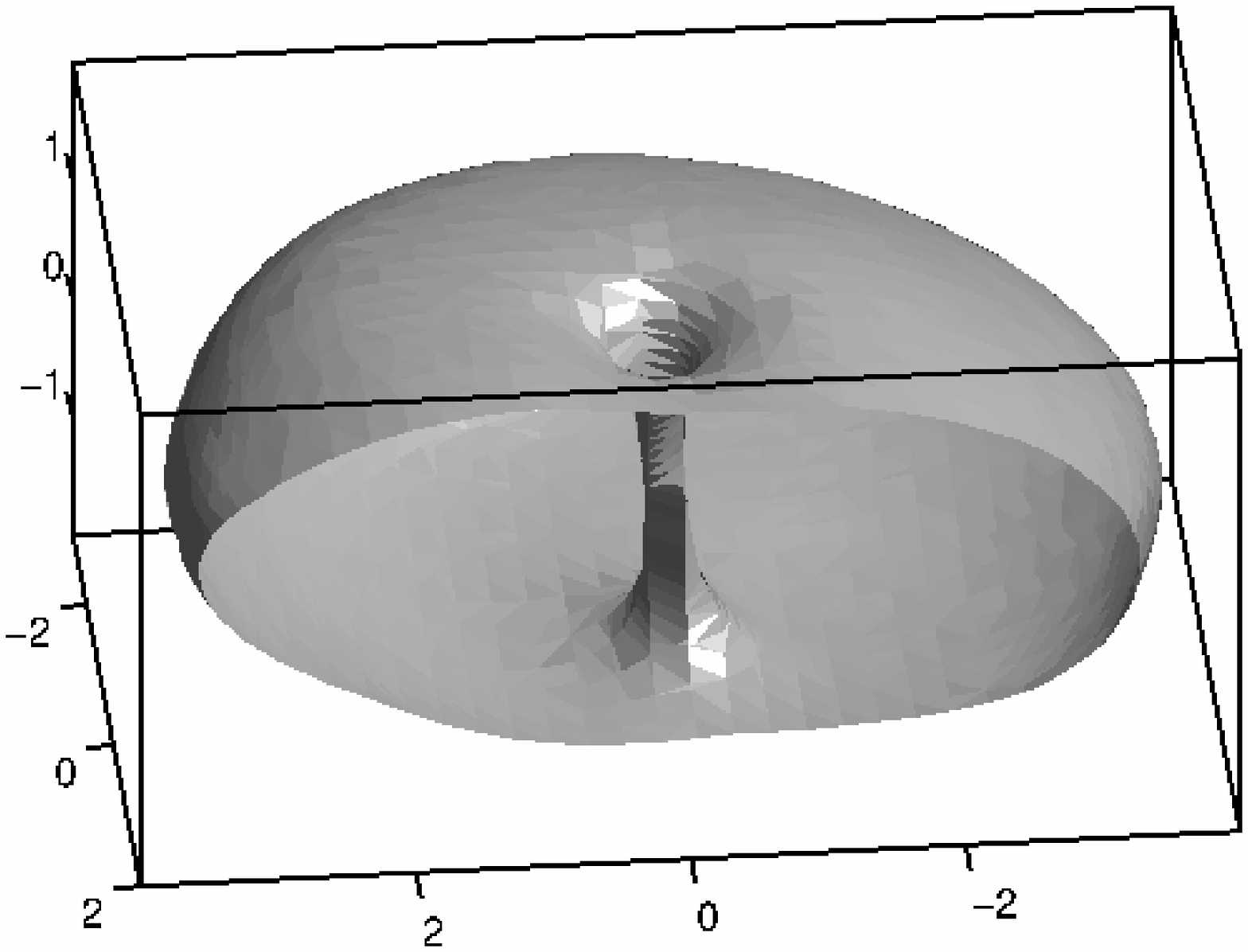}
\includegraphics[width=5cm,angle=0]{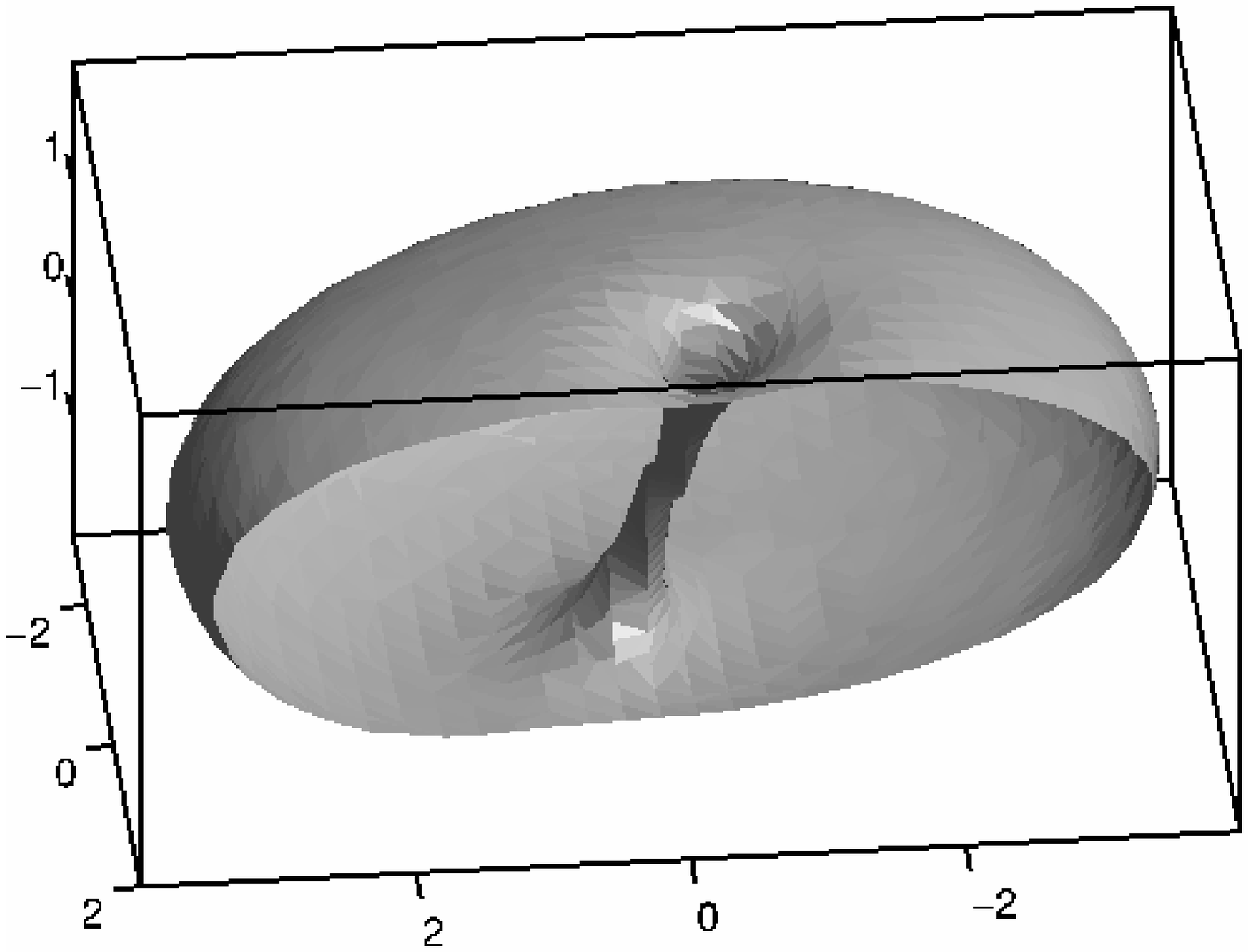}
\includegraphics[width=5cm,angle=0]{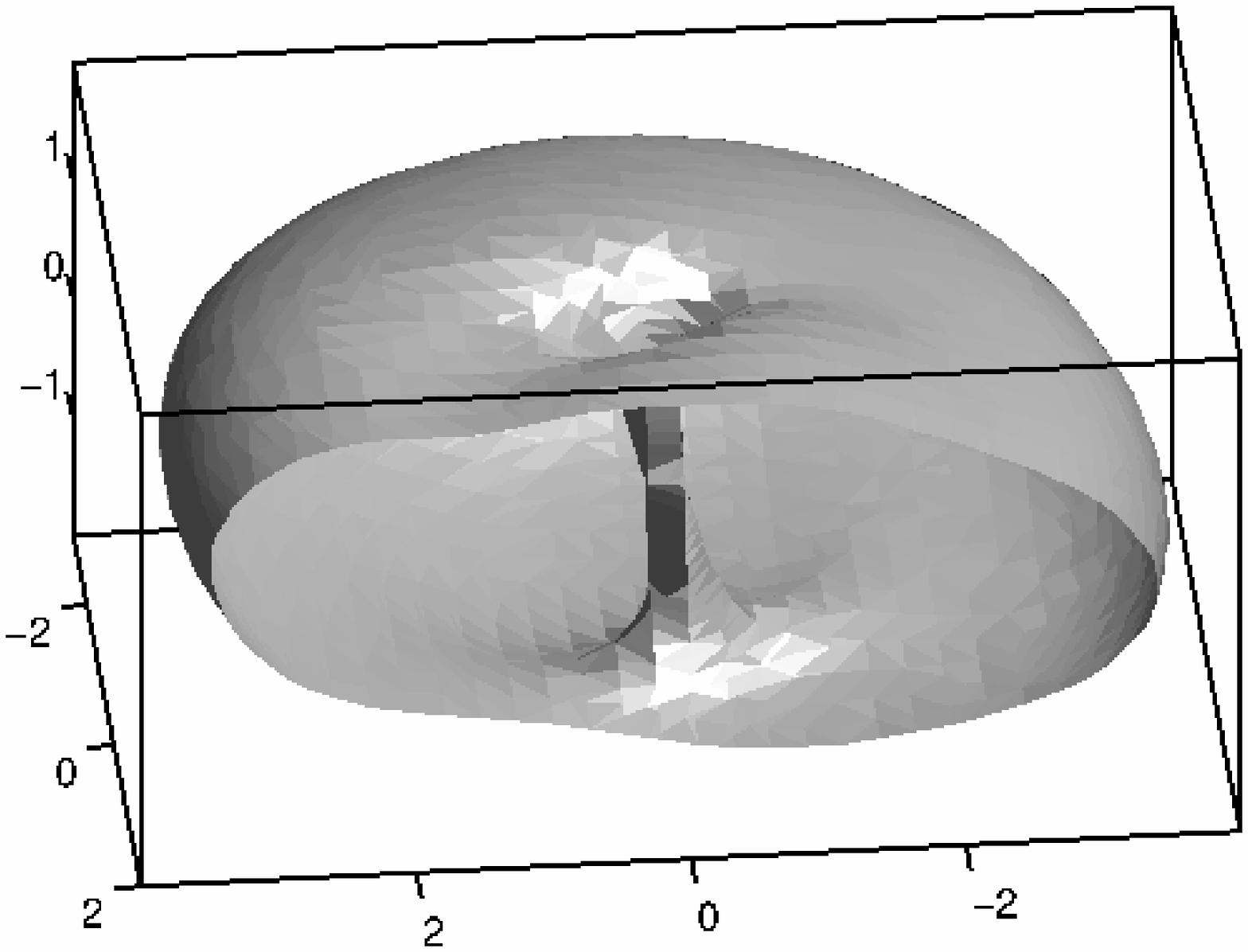}}
\caption{Surfaces of constant density of the condensate. Results 
from the solution of the Gross-Pitaevskii equation showing 
vortex bending and twisting, and condensate quadrupole oscillations 
in the scissors mode.  A
vertical cut through the surface reveals the vortex core shape.
The images correspond to the parameters $C=1000$,
$\lambda=\sqrt{7}$, and  angle $\theta_0 =10^{\circ}$.
The figure shows a sequence of frames at regular intervals.
Top row, left to right  $t=0.0, 1.3,    2.6 $, then bottom 
row, left to right $t=3.9  ,  5.2  $ and 
$t=6.5$ in units of $\omega_0^{-1}$. The vertical $z$-axis
points in the direction of the axisymmetric  trap axis. The unit 
of length on the figures is $(\hbar / 2 m_a \omega_0)^{1/2}$.}
\label{figure_2}
\end{figure}

Consider a dilute system of $N_0$ atoms, each of mass $m_a$, trapped by an 
external potential $V_{\rm ext}(\bi{x},t)$ and interacting weakly
through the two-body potential $V(\bi{x},\bi{x}')$.
At low temperatures and densities, the atom-atom interaction
can be represented perturbatively by the $s$-wave 
pseudopotential:  $V(\bi {x},\bi{x}')=  
(4\pi\hbar^{2}a_s/m_a)\delta ^{(3)}(\bi{x}
 - \bi{x}')$, and 
 $a_s$ is  the $s$-wave scattering length. The dynamics 
follow from the Hartree variational principle:
\begin{equation}
\delta \int_{t_1}^{t_2}  dt \int d^3 \bi{x} \ \ \psi^*[ H_0 +
\textstyle{1 \over 2} g \psi^* \psi-i\hbar\partial_t ]\psi =0
\label{vp}
\end{equation}
where $g=(4\pi \hbar^2 / m_a) N_0 a_s$, 
$H_0 = -(\hbar^2/2m_a)\nabla^2+V_{\rm ext}-\mu$, and 
the chemical potential $\mu$ plays the role of a Lagrange multiplier.
The condensate and its excitations 
can be described by the linear response ansatz:
\begin{equation}
\psi(\bi{x},t)  =   a_0(t) \phi (\bi{ x}) 
  +  \sum_{j>0} \left[ a_j(t)  {u}_j(\bi{x}) e^{-i\omega_j t}
+ a^*_j(t) {v}^*_j (\bi{x}) e^{+i\omega_jt}\right] 
\label{ansatz}
\end{equation}
where $\phi$ represents the highly-occupied condensate; that is,
$ | a_0| \approx \sqrt{N} \gg |a_j|$, $j>0$.
From the variation $\delta \phi^*$, and 
linear expansion  in the small parameters $a_j,a_j^*$ taken as 
constant, the stationary Gross-Pitaevskii equation and 
Bogoliubov equations follow:
\begin{equation}
H_0 \phi +g |\phi|^2\phi =0
\label{gpe}
\end{equation}
with $ (\phi, \phi) =1$  and $ \int d^3{\bi
x}\; f({\bi x})^* g({\bi x}) \equiv (f,g)$. 
The Bogoliubov modes are solutions of the
coupled linear equations:
\begin{eqnarray}
(H_0+2g |\phi|^2) u_j +g \phi^2\  v_j & = & +\hbar\omega_j u_j 
\label{bdg1} \\
(H_0+2g |\phi|^2)v_j +g \phi^{*2} u_j & = & -\hbar\omega_j v_j 
\label{bdg2}
\end{eqnarray}
Time-reversal symmetry of equations (\ref{bdg1},\ref{bdg2}) is reflected in
the fact that every set of solutions $\{\omega_j,u_{j},v_{j}\}$
has a corresponding set $\{-\omega_j,v^{\ast}_{j},u^{\ast}_{j}\}$ 
and the  normalisation can be chosen conveniently, such that:
$ (u_i,u_j)-(v_i,v_j) =\delta_{ij}$.

\begin{figure}
\centering
\includegraphics[width=11cm,angle=0,clip=true]{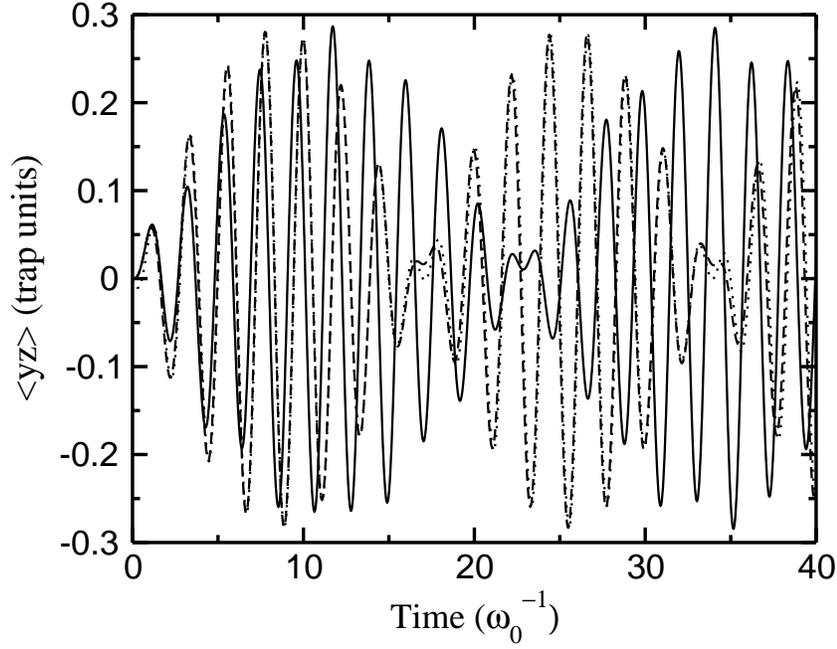}
\caption{Quadrupole oscillations in the vertical plane. Results for
 $\langle yz  \rangle (t)\equiv (2/N_0) Q_{yz}(t) $ for $\theta_{0}=4^{\circ}$, $C=1000$ and $
\lambda=\sqrt{7}$ are shown. Gross-Pitaevskii  simulation
(solid line), coherent hydrodynamic state model (dotted line), and the
 classical model (dashed line). The quantum frequency shift leads to a slower precessional 
motion than predicted by classical theory.}
\label{figure_3}
\end{figure}

\begin{figure}
\centering
\includegraphics[clip=true,width=10cm,angle=0]{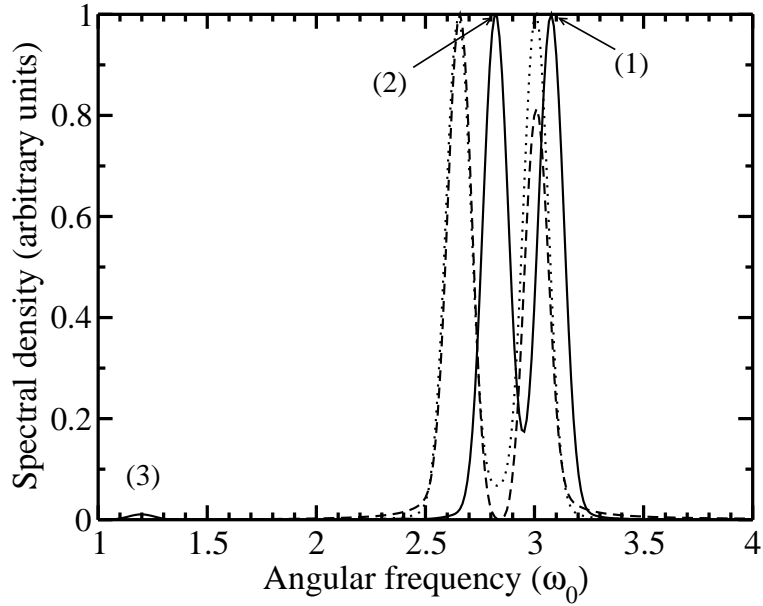}
\caption{Spectral analysis $P_{yz}(\omega)$ (equation \ref{pqij}) 
of the  vertical-plane quadrupole
 $\langle yz  \rangle (t)$ shown in figure \ref{figure_3}.
  The parameters are  $\theta_0=4^{\circ}$, $C=1000, 
\lambda =\sqrt{7}$.
 Results of Gross-Pitaevskii  simulation (solid line)
from equation (\ref{qqij}), coherent hydrodynamic states (dotted line) from
equation  (\ref{qhij}). Also shown is 
the  classical model data (dashed line) obtained from
equations  (\ref{pqij}) and (\ref{qclass}). The higher 
frequency 
peak (1) corresponds to the  quadrupole  $(h;0,1,1)/(i;0,2,1)$ 
and  the lower frequency (2) to the quadrupole $(h;1,-1,1)/(i;1,0,1)$.  } 
\label{figure_4}
\end{figure}

\subsection{Superfluid dynamics}  The equivalent fluid
motion is expressed through the coupled differential equations for 
mass and momentum transport \cite{flow_review}. 
The mass density $\rho$ and momentum current density
$\mbox{\boldmath $J$}$ are defined as, $\
\rho\equiv m_a\psi^*\psi$ and $J_k \equiv (\hbar/2i)(\psi^*
\partial_k \psi -\psi \partial_k \psi^*)$, where the index $k$
denotes the $k$th vector component, and repeated index summation
convention is used. The fluid velocity is defined by $v_k\equiv
J_k/\rho$, or equivalently in terms of the phase, $\chi$, of the
wavefunction, $v_k \equiv (\hbar/m_a)\partial_k \chi$. Then mass
and momentum  conservation are expressed by:
\begin{eqnarray}
&\partial_t \rho + \partial_j J_j =0~ \\
\label{mass}
&\partial_t J_k + \partial_j T_{jk}+\rho \partial_k (V_{\rm ext}/m_a)=0~,
\label{mom}
\end{eqnarray}
where the momentum flux density tensor is defined;
\begin{equation}
T_{jk} =   \rho v_j v_k +{\textstyle{1 \over 2}}\delta_{jk} g
(\rho/m_a)^2 -(\hbar/2m_a)^2 \rho
\partial_j \partial_k \ln \rho~.
\label{eq:shear}
\end{equation}
Finally, the mass and momentum  equations (6,7) combine
to give the acoustic  equation:
\begin{equation}
\partial_{tt} \rho = \partial_{i}\partial_{j} T_{ij} + {\partial_i}
 \left[ \rho {\partial_i } (V_{\rm ext}/m_a) \right]
 \label{acoustic}
\end{equation}
The pressure is defined as \cite{flow_review},
 $ p \equiv  {\textstyle{1 \over 2}} g
(\rho/m_a)^2 -(\hbar/2m_a)^2 \rho \nabla^2 \ln \rho~$. The first 
term represents classical pressure, the second term is 
the quantum pressure and is discarded in the classical (hydrodynamic) 
approximation. The static ($J_k=0$)
equilibrium density  in the 
hydrodynamic limit  follows from (\ref{mom}):
\begin{equation}
\rho^{(h)}_0 ({\bi x}) = (m_a/g)\left[ \mu-V_{\rm ext}({\bi x},0)
\right] \label{hstatic}
\end{equation}
For small flow and density fluctuations about
equilibrium, the kinetic energy and quantum pressure 
terms are small, so that:~$T^{(h)}_{ij} \approx{\textstyle{1 \over 2}} g
(\rho/m_a)^2 \delta_{ij}$. Using equation (\ref{acoustic}) the
hydrodynamic acoustic  modes  given by 
 $\rho ({\bi x},t) =\rho_{0}^{(h)}({\bi x})+ \bar{\rho}^{(h)}({\bi x}) e^{-i\omega t}  $
satisfy the 
linear eigenvalue problem:
\begin{equation}
-m_a \omega^2 \bar{\rho}^{(h)} = \partial_i \left[ \left[ \mu -
V_{\rm ext} ({\bi r},0) \right] \partial_i \bar{\rho}^{(h)}
\right] \label{hydro}
\end{equation}
Finally, the relation of the torque ($\tau_i$)  by an external moment to the
rate of change of angular momentum is given by:
\begin{equation}
\tau_i (t) = \partial_t \int_{\Omega}  \ \epsilon_{ijk} x_j
J_k  \ d \Omega = - \int_{\Omega}  \epsilon_{ijk} x_j \left[
\partial_l T_{lk}+\rho \partial_k (V/m_a)\right]\ d \Omega \label{torque1}
\end{equation}
where the fluid occupies the volume $\Omega$, and
$\epsilon_{ijk}$ is the Levi-Civita symbol.
Since $T_{ij}=T_{ji}$  it follows that:
\begin{equation}
\tau_i (t) =  - \int_{\Omega}  \epsilon_{ijk}\  \rho\ x_j
\partial_k (V/m_a) d \Omega
 \label{torque2}
\end{equation}
The conservation of circulation, Kelvin's theorem, and its
quantisation means that within the fluid, only those moments above
a  critical value can transfer angular momentum
associated with vortex creation or destruction. Suppose the trap 
orientation changes with time, then the ellipsoidal potential
can be written as the quadratic form: 
$V(t) = {\textstyle{1 \over 2}} a_{ij}(t) x_i x_j $
where $a_{ij}=a_{ji}$. 
Then the  equation for superfluid motion, that is 
the absence of torque $\tau_i=0$,
is  simply a set of homogeneous linear equations 
for the quadrupole moments of the 
condensate:
\begin{equation}
a_{kl}(t)\ \epsilon_{ijk}
 \int_{\Omega}   \rho(\bi{x},t) \ x_j x_l \ d \Omega = 0 
 \label{torque3}
\end{equation}
These moments are the 
key physical parameters for the condensate motion.

\section{Methods}

\subsection{Bogoliubov-de Gennes equations}

Consider the atoms confined by  an 
oblate spheroidal (pancake-shaped) trap: 
$V_{\rm ext} ({\bi x},0) = {1 \over 2} m_a \omega_0^2
(r^2+\lambda^2 z^2)$, 
where $r$ is the radial coordinate and  $z$ the axial 
coordinate, with $\lambda>1$ the aspect ratio of the trap.
 For 
convenience we use scaled dimensionless units for length, time
and energy namely: $(\hbar/2m_a\omega_0)^{1 \over
2},\omega_0^{-1}$ and $ \hbar \omega_0$, respectively. 
We define the interaction  strength by the 
dimensionless parameter $C \equiv  8\pi N_0 a_s 
 (\hbar/2m_a\omega_0)^{-{1 \over
2}}$, so that $C \rightarrow 0$ represents the ideal gas and $C
\rightarrow \infty$, describes the hydrodynamic limit. Experiments
studying the scissors and gyroscope modes performed by the Oxford group
\cite{sis_oxford1,sis_oxford2,sis_oxford3,ho02} employ  the atom 
 ${\rm Rb}^{87}$ for which   $a_s \approx 110$ a.u.. So a typical value of
$C$ corresponding to a trap frequency
 $\omega_0=2\pi \times 110 \ 
{\rm Hz}$ and particle number $ N_0 \sim 5000 $ would be $C\sim
1000$, while $\omega_0=2\pi \times 62 \ 
{\rm Hz}$ with $N_0 \sim 19000$ is equivalent to $C \sim 2870$.
In the ideal gas limit, the 
energy of the state which we label $(i;n_r,n_{\theta},n_z)$, 
has the value:
\begin{equation}
E (i; n_r,n_{\theta},n_z) = 2n_r +|n_{\theta}| +1 + (n_z+{\textstyle{1 \over 2}})
\lambda
\end{equation}
with $n_{\theta}=0,\pm 1,\pm
2,\dots $ and the radial and axial quantum numbers are:
$ n_r, n_{z}=0,1,\dots$. The corresponding 
excitation frequencies, with respect to 
a  vortex state $ (i;0,\kappa,0)$, 
where $\kappa=0,\pm 1,\pm 2, \dots$, are given by:
\begin{equation}
\omega (i; n_r,n_{\theta},n_z) = 
2n_r + |n_{\theta}| -|\kappa| + n_z \lambda
\label{ispec}
\end{equation}

For finite $C$, the spectrum of excitations 
must be determined by numerical solution of 
equations (\ref{gpe},\ref{bdg1}) and 
(\ref{bdg2}). Separating variables gives:
\begin{equation}
\phi(r,z,\varphi) = \tilde{\phi}_{\kappa}(r,z)e^{i\kappa\varphi}
\end{equation}
so that the condensate with circulation $\kappa$ is the solution of the
equation:
\begin{equation}
- \left( {\partial^2 \over \partial r^2}
+ \frac{1}{r} \frac{\partial}{\partial r}
+ \frac{\partial^{2}}{\partial z^{2}}
- \frac{\kappa^{2}}{r^{2}} \right) \tilde{\phi}_{\kappa}
+ {\textstyle {1 \over 4}}(r^2+\lambda^2 z^2)
\tilde{\phi}_{\kappa} + C |\tilde{\phi}_{\kappa}|^{2}
\tilde{\phi}_{\kappa}= \mu \tilde{\phi}_{\kappa}
\label{condeq}
\end{equation}
We use  $m=n_{\theta}-\kappa$ to denote 
angular momentum with respect to the condensate, that is  
the helicity of the excitations, and  it is convenient to use
the labelling  $(h;q_r,m,q_z)$ appropriate 
for the hydrodynamic limit ($C \rightarrow \infty$).  Thus the 
quasiparticle amplitudes:
\begin{equation}
\fl  u_{n_r,m,n_{z}}(r,z,\varphi) \equiv
\tilde{u}_{n_r,n_z}(r,z) e^{i(m + \kappa)\varphi} \ \ \ \  {\rm and} 
\ \ \ \ 
v_{n_r,m,n_z}(r,z,\varphi) \equiv
\tilde{v}_{n_r,n_z}(r,z) e^{i(m - \kappa)\varphi}
\end{equation}
with corresponding angular frequency,
$\omega_{n_r,m,n_{z}}$,
are solutions of the eigenvalue problem
\begin{eqnarray}
{\cal{L}}(m+\kappa)\tilde{u}_{n_r,n_{z}}(r,z) +
C\ \tilde{\phi}_{\kappa}^{2}\ \ \ \tilde{v}_{n_r,n_{z}}(r,z)
&=& \omega_{n_r,m,n_{z}}\tilde{u}_{n_r,n_{z}}(r,z)  \label{dimbdg1}\\
{\cal{L}}(m-\kappa)\tilde{v}_{n_r,n_z}(r,z) +
C\ \tilde{\phi}_{\kappa}^{\ast \ 2}\ \tilde{u}_{n_r,n_{z}}(r,z)
&=& \omega_{n_r,m,n_{z}}\tilde{v}_{n_r,n_{z}}(r,z) \label{dimbdg2}
\end{eqnarray}
where
\begin{equation}
{\cal{L}}(s) \equiv
- \left( {\partial^2 \over \partial r^2}
+ \frac{1}{r} \frac{\partial}{\partial r}
+ \frac{\partial^{2}}{\partial z^{2}}
- \frac{s^{2}}{r^{2}} \right) \tilde{\phi}_{\kappa}
+ {\textstyle {1 \over 4}}(r^2+\lambda^2 z^2)
\tilde{\phi}_{\kappa} + 2C |\tilde{\phi}_{\kappa}|^{2}
\end{equation}
In our calculations, these two-dimensional equations are discretised  by
Lagrange meshes  \cite{dermot02}; 
 the radial coordinate is defined at $M$  grid points
($r_1,r_2,\dots,r_{M}$)
and the axial coordinate at $N$ points
($z_1,z_2,\dots,z_{N}$). Therefore:
\begin{equation}
\tilde{\phi}_{\kappa}(r,z) = \sum_{k=1}^{M}\sum_{l=1}^{N}
\tilde{\phi}_{\kappa}^{kl} (r_{k},z_{l}) \lambda_{k}^{-1/2} \mu_{l}^{-1/2}\
f_{k}(r)g_{l}(z)
\label{3ddvr}
\end{equation}
where $f,g$ are Lagrangian interpolating functions such that
\begin{eqnarray}
\int_{0}^{\infty} f_{i}^{\ast}(r)f_{k}(r) 2\pi r \ dr \approx
\lambda_{i}\delta_{ik}  \\
\int_{-\infty}^{\infty}g_{j}^{\ast}(z)g_{l}(z)\ dz \approx \mu_{j}\delta_{jl}
\end{eqnarray}
%Each quasiparticle pair of functions
%($\tilde{u}, \tilde{v}$) is treated in  the same way.
The Lagrange functions for the $r$-coordinate are 
chosen to be generalised Laguerre polynomials \cite{dermot02}, 
scaled to encompass the entire condensate, with typically
$M =50$ mesh points. Hermite polynomials are used
in the $z$-direction so that
\begin{equation}
g_{l}(z) = \sum_{l=0}^{N-1} \chi_{l}^{\ast} (z_{l})  \chi_{l}(z)
\end{equation}
where $\chi_{l}(z) = h_{N}^{-\frac{1}{2}} w(z)^{\frac{1}{2}} H_{l}(z)$.
 and $H_{l}(z)$ are the Hermite polynomials associated with weights
 $w(z)=e^{-z^2}$  and normalisation factor
 $h_{N} = 2^{N}\pi^{1/2}N\!$.
 A high degree of accuracy was found with only $N=30$ points.
The resultant eigenvalue problem  was
solved using Newton's method for equation (\ref{condeq}) and a standard eigenvalue
routine for equations (\ref{dimbdg1},\ref{dimbdg2}).
Convergence was established by a combination of grid scaling
and number of mesh points, so that at least six-figure accuracy was 
assured for all frequencies (see table \ref{table_1}).

\begin{figure}
\centering
\includegraphics[clip=true,width=12cm,angle=0]{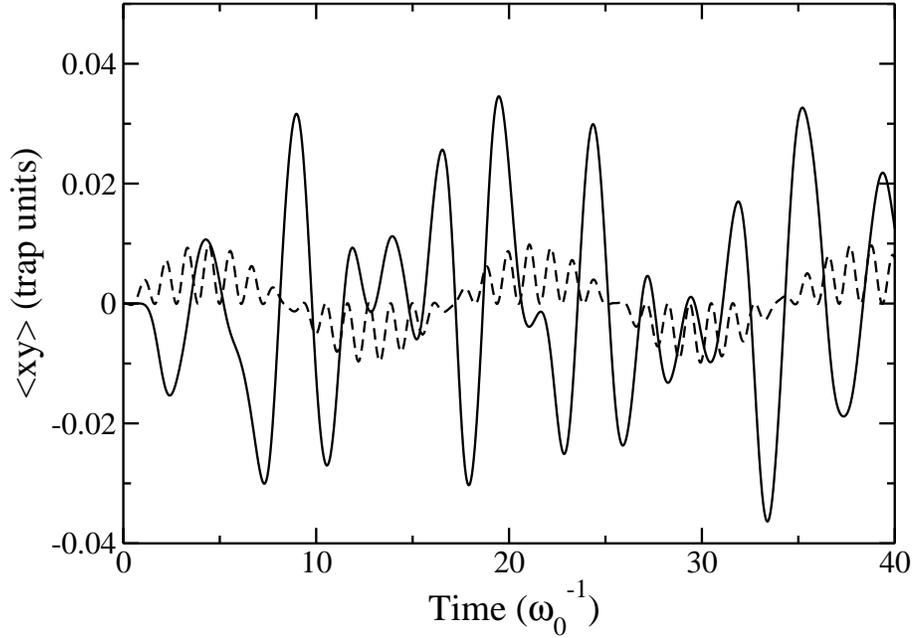}
\caption{Quadrupole variations in the horizontal plane. Results for  
$\langle xy \rangle (t) \equiv (2/N_0) Q_{xy}(t) $ for angle
$\theta_{0}=4^{\circ}$,
$C=1000$ and $\lambda=\sqrt{7}$. The full line is the 
Gross-Pitaevskii quantal simulation,  the dashed line is the 
classical model. In contrast to the results in figure \ref{figure_3}, the
two models strongly disagree.} 
 \label{figure_5}
\end{figure}

\begin{figure}
\centering
\includegraphics[clip=true,width=10cm,angle=0]{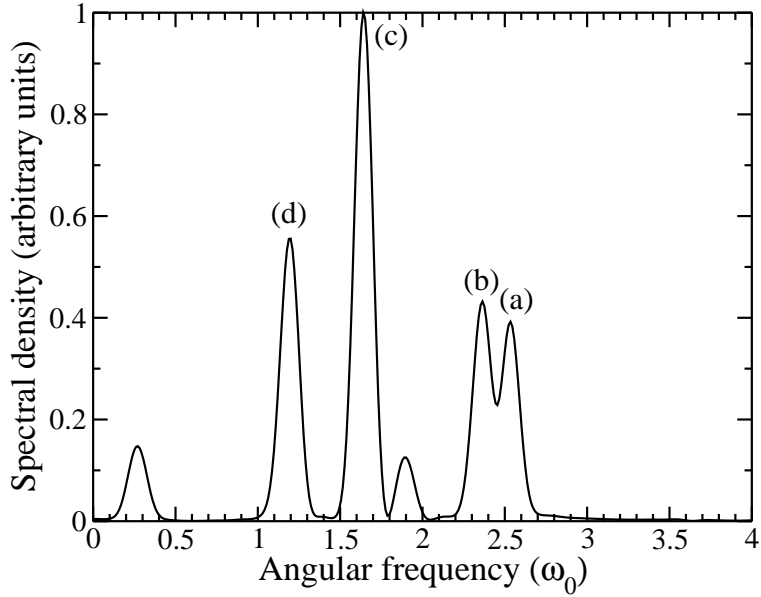}
\caption{Helical mode spectrum: the power spectrum
 $P_{xy}(\omega)$, given by equation (\ref{pqij}) 
 of the horizontal-plane quadrupole shown 
 in figure \ref{figure_5}.
 The Kelvin mode, (c) $m=+2$, is most strongly excited 
 and  leads to helical oscillations 
 of the vortex core. The mode (d) $m=-2$ is more weakly coupled, 
and the lines (a,b) associated with 
 $m=\pm 1$ approximately cancel.}
\label{figure_6}
\end{figure}

\subsection{ Time-dependent Gross-Pitaevskii equation}

The time-dependent Gross-Pitaevskii
equation, follows from taking 
arbitrary variation of $\psi^*$ in (\ref{vp}):
\begin{equation}
 \left[ H_0 + g |\psi|^2 -i\hbar\partial_t \right]\psi =0
\label{nlse}
\end{equation}
The direct numerical solution of this equation, without linearisation, 
 in combination with spectral analysis can also
be used to determine the frequency spectrum \cite{dermot02}
and density fluctuations of the collective excitations.
This can be done efficiently and accurately using spectral 
methods. The ground state of the system is found
by evolving equation (\ref{nlse}) in imaginary time
using the split-step Fast-Fourier transform propagator \cite{dermot02}.
A simple arbitrary trial function is used as the initial 
state. At each time step in the evolution 
the fluid circulation is imposed by a applying 
a phase gradient corresponding to $\kappa=+1$ until the 
excited states diffuse from the system and the 
density and chemical potential
stabilise. In practice this normally takes a few trap periods at most.
Having found the condensate in this manner, the system 
is then allowed to evolve
in real time under any external perturbations using the same numerical
method.  

For example a sudden  rotation of the trap by a small angle will disturb 
the steady state flow of the condensate and create a variety 
of small amplitude excitations. 
Under these conditions, the linearisation (\ref{ansatz}) 
is valid and the mode frequencies and densities correspond those 
of the  Bogoliubov-de Gennes equations (\ref{bdg1},\ref{bdg2}). 
It follows that the expectation value of any operator
will exhibit beats between the various
modes. Since a simple rotation does not induce centre-of-mass (dipole)
motion, the lowest-order distortion will arise in the quadrupole moments:
\begin{equation}
Q^q_{ij}(t) \equiv  \int \psi^*({\bi x},t)\; x_i x_j \; \psi({\bi x},t) \ d{\bi x}
\label{qqij}
\end{equation}
where $\psi({\bi x},t)$ is the solution of equation (\ref{nlse}).
The condensate quadrupole 
moments in the vertical and horizontal planes will give 
 information on the vortex bending and twisting, respectively.
We present results for the scaled moments:
$ \langle yz \rangle \equiv (2/N_0) Q_{yz} (t) $, in figure 
(\ref{figure_3}), and 
$ \langle xy \rangle  \equiv (2/N_0) Q_{xy} (t) $, 
in figure (\ref{figure_5}),
and the corresponding power spectral densities:
\begin{equation}
P_{ij} (\omega) \equiv \left| \int_0^T e^{-i\omega t}\; Q_{ij}(t)\ dt \right| ^2
\label{pqij}
\end{equation}
in figures (\ref{figure_4}) and (\ref{figure_6}), respectively.

\subsection{ Coherent hydrodynamic states}

The  hydrostatic spectrum, equation (\ref{hydro}), also follows
from a simplified version of the  the Bogoliubov-de Gennes
 equations   \cite{ohberg}
 when  quantum pressure has been neglected:
\begin{eqnarray}
q^2 \left[ -\nabla^2+q^{-1}(\nabla^2 q) \right] h^+_{j}
  -{\textstyle{1 \over 2}} \omega_j^2 h^+_j =0 \\
\label{h1}
\left[ -\nabla^2+q^{-1}(\nabla^2 q) \right] (q^2 h^-_j)
  -{\textstyle{1 \over 2}} \omega_j^2 h^-_j =0
\label{h2}
\end{eqnarray}
where $h^{\pm}_j=u_{j}\pm v_{j}$, and $q^2$ is the vortex-free hydrostatic
density given by 
\begin{equation}
\fl q^2(x,y,z) 
= \left\{ 
\begin{array}{ll}
\left[ \mu -{\textstyle{1 \over 4}} \left(
x^2+y^2+\lambda^2z^2\right) \right] & \textrm{if $ 
\mu -{\textstyle{1 \over 4}}\left(
x^2+y^2+\lambda^2z^2\right) \geq 0  $} \\ 
 0  & \textrm{if $ 
\mu -{\textstyle{1 \over 4}}\left(
x^2+y^2+\lambda^2z^2\right) < 0  $} \\ 
\end{array}
\right.
\label{tf}
\end{equation}
These equations can be solved in closed analytic form using
 series expansions \cite{ohberg} to yield expressions for the
 quasiparticle amplitudes. The scissors mode
quadrupole ($\kappa=0,m=\pm1$) for $\lambda > 1$,
 with $\bar{\rho}^{(h)}({\bi x})  \propto xz $,
first proposed by Gu\'ery-Odelin and Stringari
\cite{sis_string}  and experimentally observed
by Marag\'o  et al. \cite{sis_oxford1} has the frequency:
\begin{equation}
\Omega_{s} = \sqrt{1+\lambda^2}.
\label{scissors}
\end{equation}
Quadrupoles corresponding to $\kappa=0,m=\pm2$ are degenerate with
frequency:
$\Omega_{xy}= \sqrt{2}$.

\begin{table}
\caption{Table of 
mode excitation  frequencies for $\lambda=\sqrt{7}$
as a function of the interaction strength:
 $C = 8\pi N_0 a_s  (\hbar/2m_a\omega_0)^{-{1 \over 2}}$.
 The full spectrum is shown in figure \ref{figure_7}.
The selected modes in the table are are: (1)\ \ ($h$;0,1,1); 
 (3)\ \ ($h$; 0,-1,1);
 (c)\ \ ($h$; 0,2,0);
and  (d)\ \  ($h$; 0,-2,0).
The first entry in each column is the result from the
Bogoliubov-de Gennes equations (3--5).
The numbers in brackets correspond to the time-dependent linear response
method of calculation.  The Bogoliubov-de Gennes results for the 
frequency of precession are denoted by $\omega_p$. All 
angular frequencies are in units of $\omega_0$.}
\begin{indented}
\item[]
\begin{tabular}{|l|c|cccccccccccc|}\hline
$C$   & $\omega_p$ & \multicolumn{2}{c}{ (1)}
&&\multicolumn{2}{c}{  (3)}&&\multicolumn{2}{c}{(c)}&&\multicolumn{2}{c}{(d)} &\\ \hline
0     &  0.000 & 3.646 &(3.646)&& 1.646   &(1.645)&& 2.000 &(2.001) && 0.000  &(0.000)&\\
50    &  0.075 & 3.488 &(3.495)&& 1.625   &(1.627)&& 1.946 &(1.896) && 0.522  &(0.516)&\\
100   &  0.110 & 3.401 &(3.398)&& 1.590   &(1.591)&& 1.889 &(1.846) && 0.738  &(0.730)&\\
250   &  0.145 & 3.267 &(3.257)&& 1.483   &(1.491)&& 1.787 &(1.780) && 0.984  &(1.016)&\\
500   &  0.149 & 3.169 &(3.165)&& 1.354   &(1.370)&& 1.708 &(1.744) && 0.111  &(1.105)&\\
1000  &  0.137 & 3.085 &(3.079)&& 1.201   &(1.194)&& 1.640 &(1.644) && 1.197  &(1.195)&\\ \hline
\end{tabular}
\end{indented}
\label{table_1}
\end{table}
The scissors mode, selected by sudden rotation $\theta_0$,
introduces a perturbation:
\begin{equation}
V' \approx m_a\omega_0^2 \theta_0(1-\lambda^2) xz \label{sciss}
\end{equation}
In the absence of a vortex ($\kappa=0$), coupling to the
 degenerate $m=\pm 1$ states
is equal and opposite, and no helicity arises.
If a vortex is present ($\kappa=1$), the  coupling
is asymmetric since the state $m=+|m| $  rotates with the 
condensate flow at frequency $\omega^+$,
whereas the mode $m=-| m| $ is counter-rotating at
frequency $\omega^-$, see figure
\ref{figure_1}. 
 The frequency difference  was
calculated  analytically 
using a hydrodynamic
model  \cite{zamb_stri} with the result that
\begin{equation}
\omega^+ - \omega^-=\frac{2\langle L_z \rangle}{m_a \langle
r^{2}+2z^2\rangle} =\frac{\langle L_z
\rangle}{I_{xx}}=\frac{\langle L_z \rangle}{I_{yy}}
\end{equation}
where $I_{xx}$ is the principal  inertia moment about a horizontal axis.
The result also agrees with the
result of Svidzinsky and Fetter \cite{vort_exfq} which treats the
vortex core as a perturbation of the hydrostatic
Thomas-Fermi density. The same result can be found and understood  
using a classical model of the motion (section \ref{sec:classical}).

To obtain the amplitude of precession and nutation,
the populations of the excited modes are required. 
Since the tilting  is small  and sudden, we can 
use diabatic perturbation theory.
A small sudden rotation by an angle
$\theta_0$ about the $y$-axis, with  generator $L_y=(2i)^{-1}(L_+-L_-)$, gives 
rise to the function:
\begin{equation}
\psi_h (t=0) \approx \exp \left[ - i \hbar^{-1}\theta_0 L_{y} \right]
\phi_h
\end{equation}
The amplitudes are found by projection  to the modes
 in the  rotated frame of reference. From equation (\ref{ansatz}):
$$
\psi_h(\bi{x},t) e^{i\mu t}  \approx 
 \sqrt{N_0} \ \phi_h (\bi{x},t)+\sum_{j} \left[
a_j(t)  u_j(\bi{x}) e^{-i\omega_j t}+ a^*_j(t)
 v_j^*(\bi{x}) e^{+i\omega_j t} \right]
$$
and the coefficients $a_j(t)$ will be time-independent. 
Projection by the hydrodynamic state
 $h^-_j \equiv u_j-v_j$ gives the time-independent amplitudes $a_j$
\begin{equation}
  (h^-_j,h^+_j)\  a_j  \approx  (h^-_j,\exp\left[-{\textstyle {1 \over 2}} 
 \hbar^{-1}\theta_0 (L_+-L_-) \right]\phi_h)
 \label{mat}
\end{equation}
 The rotation mixes the initial state 
with $m=\pm 1,\pm 2$, to order $\theta_0$, and $\theta_0^2$, 
respectively. We have overlooked the density distortion of the vortex, though 
included the angular momentum. 
Since the  vortex core distorts the
density  over a comparatively small volume of the
condensate, then to a first approximation the density is the
axisymmetric Thomas-Fermi distribution, equation (\ref{tf}).
Therefore, as a simple first-order approximation we  take
\[
\phi_h = C^{-{1 \over 2}} \ q(r,z)  \ e^{i\kappa \varphi}
\]
where $q^2$ is given by equation (\ref{tf}). 
The time-independent equations (30,31) can be used 
to calculate the quasiparticle functions, and the transition amplitudes,
$a_j$ given by (\ref{mat}), can be extracted analytically
as shown in  \cite{halvor_coherent}.
In this approximation, the excitation spectrum is the same as the vortex free 
case, and does not include the degeneracy splitting  arising
in higher-order perturbation theory \cite{zamb_stri,vort_exfq}. 
The  quadrupole excitations in the vertical and 
horizontal plane  are defined as:
\begin{eqnarray}
 Q^h_{yz}(t) \equiv  \int \psi_h^*({\bi x},t)\; y z\;
\psi_h({\bi x},t) \ d{\bi x}  \\
Q^h_{xy}(t) \equiv  \int \psi_h^*({\bi x},t)\; xy\;
\psi_h({\bi x},t) \ d{\bi x}
\label{qhij}
\end{eqnarray}
and can be  calculated analytically.
The results for this coherent hydrodynamic model
are presented  in figures \ref{figure_3} and \ref{figure_4}.

\subsection{ Classical model}\label{sec:classical}
We use the convention of Goldstein \cite{goldstein} for the Euler
angles $(\phi,\theta,\psi)$,
describing  the orientation of a set of  rotating axes
$ (x',y',z')$ with respect to a 
space fixed frame (figure \ref{figure_1}).
 The principal moments of
inertia of the vortex-free density, equation (\ref{tf}), are:
\begin{equation}
\fl
I_{xx}  =  {(\lambda^2+1) \over 3\lambda^2}I_s  \ \ \ \ {\rm and}  \ \ \
I_{zz}  =   {\textstyle {2 \over 3}}I_s \qquad{\rm where}\qquad
I_s  = {\textstyle {3 \over 7}}  m_a N_0 \left[ 15 \lambda N_0 g \over
4 \pi m_a \omega_0^2 \right]^{2 \over 5}
\end{equation}
In general the
Lagrangian density, ${\cal L}_a $,
is given by \cite{neg88}:
\begin{equation}
\mathcal{L}_a(\rho,\phi_v,\dot{\rho},\nabla\phi_v,\nabla \rho;{\bi
x};t)=
 \phi_v \dot{\rho}
- {\textstyle{1 \over 2}} \rho  (\nabla\phi_v)^2
 -\rho\ U  -{\hbar^2 \over 8m_a^2 \rho} (\nabla \rho)^2
\end{equation}
where the potential per unit mass is
 $U=(V_{\rm ext}/m_a) + (g \rho/2m_a^2)$ and  $\bi{v} = \nabla \phi_v$. Discarding  the last
 term gives the classical interacting fluid
\cite{oden} with the corresponding Hamiltonian density:
\begin{equation}
{\cal H}_c = {\textstyle{1 \over 2}} \rho  \bi{v}^2 +{g\rho \over
2m_a^2}+{\textstyle{1 \over 2}} \rho \omega_0^2
(x^2+y^2+\lambda^2z^2)
\end{equation}
 At equilibrium the condensate has a hydrostatic (vortex-free) 
density  and angular momentum $N_0\hbar
\kappa$. A small change of orientation of the trap ($\delta\theta$) creates
 density and velocity changes, $\tilde{\rho},
\tilde{\bi{v}}$:
\begin{equation}
\partial_t \tilde{\bi{v}} +(g/m_a) \nabla \tilde{\rho}=0  \ \ \  ,\ \ \ \ \  \partial_t
\rho+ \nabla \cdot (\rho_0 \nabla \tilde{\bi{v}})=0\ \ \
\end{equation}
In the body-fixed frame, and for quadrupole ($xz$) perturbations
only, this leads to an effective potential
\begin{equation}
\delta V(\theta)= {\textstyle {1 \over 2}}  \omega_0^2
(1+\lambda^2) \rho_0 (x^2+z^2)(\delta\theta)^2
\end{equation}
Then the Lagrangian in the space-fixed frame has the  form, for
small $\theta$:
\begin{equation}
\fl L(\theta,\dot{\varphi},\dot{\theta},\dot{\psi},t)=
 {\textstyle  {1 \over 2}} I_{xx}
\sin^2\theta\  \dot{\psi}^2 +{\textstyle  {1 \over 2}} I_{xx} \dot{\theta}^2+
{\textstyle {1 \over 2}}I_{zz} (\dot{\psi}+\dot{\varphi} \cos\theta)^2-
{\textstyle {1 \over 2}} I_{xx} \Omega_s^2 \theta^2
\end{equation}
From this Lagrangian, Hamilton's equations are
\begin{eqnarray}
\dot{\varphi}  =  {( p_{\varphi}-p_{\psi}\cos \theta ) \over I_{xx} \sin^2\theta}
&
\dot{\theta}  =  {p_{\theta} \over I_{xx} } \\
\dot{\psi}    =  {p_{\psi} \over I_{zz}} - {(p_{\varphi}-p_{\psi}\cos \theta)
\cos\theta \over I_{xx} \sin^2\theta }  &
\dot{p}_{\varphi}      =  0 \\
\dot{p}_{\theta}    =   {(p_{\varphi}-p_{\psi}\cos \theta
)(p_{\varphi}\cos\theta-p_{\psi}) \over I_{xx} \sin^3\theta }-
I_{xx}\Omega^2_s\theta \ \ \ \ \ \ & \dot{p}_{\psi}      =   0
\label{ham}
\end{eqnarray}
In anticipation of the numerical results, consider possible analytic
solutions of the equations for very small angular displacements $\theta$.
The top is aligned vertically ($\theta=0$) and
spinning with angular momentum $J_0$, thus $p_{\psi}=J_0$ and $p_{\theta}=0$.
The possible singularity at $\theta=0$ in the equation for $\dot{\varphi}$
can be avoided if $p_{\varphi}-p_{\psi}\cos \theta \rightarrow 0 $ as
$\theta \rightarrow 0$. A steady azimuthal precessional,
$\ddot{\varphi}=0$, independent of  $\theta(t)$,
arises from the case $p_{\psi}=p_{\varphi}=J_0 $
in which case, for  $\cos\theta \approx 1 -{\textstyle{1 \over 2}}\theta^2$,
  we have:
$$
\dot{\varphi} \approx { J_0 \over 2 I_{xx}}
$$
Under the same approximations,
 and the fact that  $\dot{\theta}=p_{\theta}/I_{xx}$:
$$
\dot{p}_{\theta} \approx - \left( {J_0^2 \over I_{xx} } +I_{xx}
\Omega_s^2\right) \theta \ \ \ \ \ \ \ \ddot{\theta}=-\left({J_0^2
\over I^2_{xx}} + \Omega_s^2\right)\theta
$$
Therefore the vortex motion is described by
\begin{equation}
\theta \approx \theta_0 \sin \sqrt{\Omega_s^2+(J_0/I_{xx})^2}.\ t
\ \ \ \ \ \ \ \ \ \ \ \ \ \varphi \approx  { J_0 \over 2 I_{xx}}t
\label{precess}
\end{equation}
which describes a steady azimuthal precession, frequency 
$\omega_p= J_0 /2I_{xx}$,  combined with sinusoidal nutation at
a  frequency slightly higher than the vortex-free scissors 
oscillation. The precessional motion splits the frequencies. It 
is equivalent 
to a steady background rotation of the condensate.
In analogy to a Doppler shift,
internal flow  around the $z$-axis associated with the quantum number $\pm m$ would be
shifted so that  $\omega^+-\omega^- \equiv 2\omega_p = J_0 /  I_{xx}$.
This  splitting, understood in terms of classical precessional motion, is
identical to the
quantum expression \cite{vort_exfq,zamb_stri}.

Initially the condensate has a pure rotation about the
$z'$-axis so that $\dot{\psi} = J_0/I_3$, and the
condensate is impulsively
tilted by an angle $\theta_0$.
The Hamilton equations are   solved   numerically, with
the initial conditions, $t=0$:
\begin{eqnarray}
\theta    =   \theta_0  \ \ \ \ \ \  & \varphi     =    0  \\   \nonumber
\psi     =    0         &     p_{\varphi}   =    J_0 \cos \theta_0\\ \nonumber
p_{\theta}  =   0    & p_{\psi}  =  J_0 \nonumber
\label{init}
\end{eqnarray}
The time dependence of the products of inertia is governed by the
orientation of the axis of the top and thus will oscillate in
time. The transformations between the laboratory and fixed-frame
quadrupole  gives
\begin{eqnarray}
Q^c_{xz}(t)  & \equiv {\textstyle {1
\over 2}}(I_{xx}-I_{zz}) \sin2\theta\ \cos\varphi \\
Q^c_{yz}(t)  & \equiv {\textstyle {1
\over 2}}(I_{xx}-I_{zz}) \sin2\theta\ \sin\varphi \\
\label{qclass}
Q^c_{xy}(t) &\equiv {\textstyle {1 \over
2}}(I_{xx}-I_{zz}) \sin 2\varphi\ \sin^2\theta
\end{eqnarray}
Since the inertia moments, $I_{xx}$ and $I_{zz}$, are constant,
the time dependence is entirely governed by $\phi(t)$ and $\theta(t)$.
These functions are determined by numerical integration of equations (\ref{ham})
with initial conditions (\ref{init}). The physical meaning of the
quadrupole  is clear if one visualises a rigid vortex motion which nutates
and precesses. At the instant the vortex is aligned
along the $y=0$ plane, then $\langle yz\rangle=0$. This happens twice 
each precession cycle. During the slow precession, the 
vortex line nutates rapidly between the 1st and 2nd 
quadrants of the $yz$-plane which leads to a change of sign in $Q_{xz}$ 
each half-cycle of the scissors period.
The amplitude of the 
vertical plane ($yz$) quadrupole  is more easily observable 
by experiment  \cite{ho02} since it is of order $\theta_0$ 
(figure \ref{figure_3}).
The horizontal plane motion, on the other hand,  is proportional to 
$\theta_0^2$ (figure \ref{figure_5}).

\begin{figure}[t]
{\centering
\includegraphics[clip=true,width=13cm,angle=0]{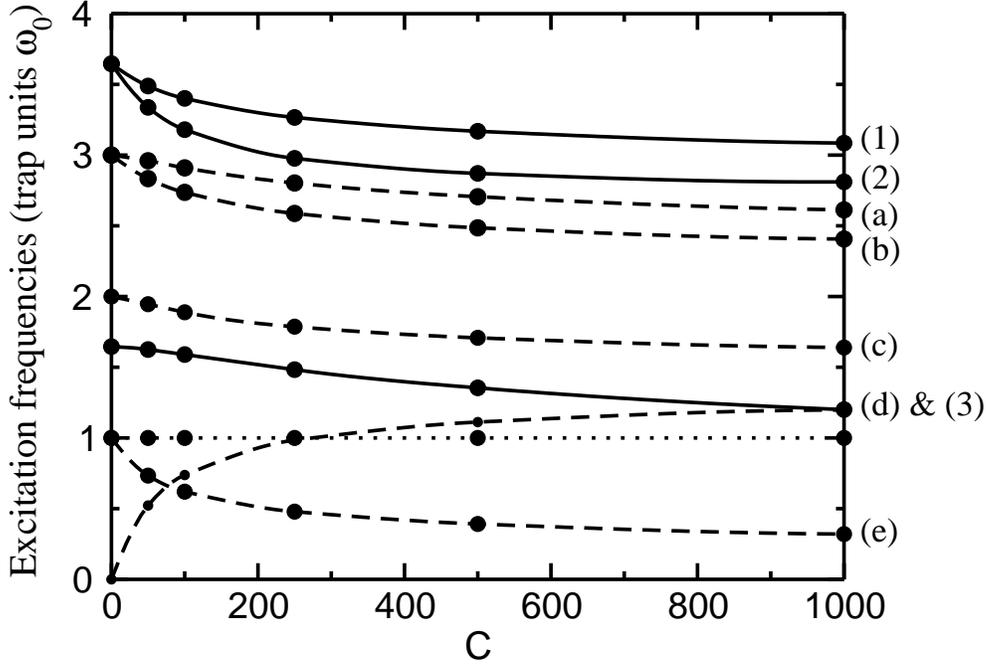}
\caption{ The excitation spectrum of the vortex condensate
as a function of interaction strength
$C =  8\pi N_0 a_s  (\hbar/2m_a\omega_0)^{-{1 \over 2}}$
. The solid lines labelled (1), (2) and 
(3) 
 correspond to those found
 in the power spectrum of $\langle yz \rangle$.  The dashed lines 
 (a),(b) , (c) and (d) are prominent in the 
 quadrupole $\langle xy\rangle$. 
The centre-of-mass dipole  mode (dotted line) is included for completeness
as well as the negative frequency mode $(h;0,-1,0)$; line (e).
The limit $C \rightarrow 0$ corresponds to the ideal gas modes
$(i;n_r,n_{\theta},n_z)$ with  frequencies given by equation (\ref{ispec}).
Values of selected mode frequencies are given in table \ref{table_1}. 
  }
\label{figure_7}
}
\end{figure}

\section{Results and Discussion}

\subsection{Excitation modes and spectrum}

Our numerical studies focus on a singly-quantised vortex ($\kappa=1$) 
condensate in a pancake shaped trap $\lambda=\sqrt{7}$, for 
detailed theoretical discussion,  and 
$\lambda = \sqrt{8} \approx 2.83$ for comparison with experiment. 
The frequencies of the 
quadrupole moments can be obtained from the excitation spectrum
of the normal modes of the system. 
The  spectra obtained for $\kappa=1$ and $0 < C < 1000$ are shown in
 figure \ref{figure_7}, and given in table \ref{table_1}.
For  $C\gg 1000$ the variation in excitation spectrum is slow and smooth, and tends 
towards asymptotic limits.
Consider  $C=1000$ with an equivalent chemical potential for the hydrostatic 
density in the absence of 
a vortex: $\mu_{\rm TF} = (15\lambda C/64\pi)^{2 \over 5} \approx 8.282 \gg 1 $.
While this lies well within the 
region of validity of the hydrodynamic approximation, in the sense that
$ \mu \gg \omega_0$,  
the trap energy and vortex core inhomogeneity are still
 significant and
one would expect quantal deviations from the 
hydrodynamic model.  
Prior to analysing the motion in detail, 
an example of the density fluctuations 
produced by our simulations is 
shown in figure \ref{figure_2}.
 The time-dependent Gross-Pitaevskii equation (\ref{nlse})  is solved 
in imaginary time with $\kappa=1$ to generate the initial 
state of the condensate. The trap  is then suddenly tilted 
by a small angle, $\theta=10^{\circ}$, about the $y$ axis and 
the evolution of the condensate follows the 
time-dependent Gross-Pitaevskii equation.
Surfaces of constant density of the condensate are shown in figure
\ref{figure_2} in the tilted space-fixed frame of reference. A
vertical cut through the surface reveals the vortex core structure
over a single trap period. The first image  $t=0$ is the initial 
configuration, the subsequent images are taken at regular intervals,
$t=1.3,    2.6  ,  3.9  ,  5.2  $ and 
$t=6.5$, respectively, in units of $\omega_0^{-1}$.
The scissors oscillation of the condensate corresponds to nutation, 
that is quadrupole oscillation in the vertical plane. This creates 
flow past the vortex producing a Magnus force which results in precession. 
However when combined with the inhomogeneity  in the condensate, the force
gives rise to vortex bending and twisting  into 
helical modes of oscillation. 
The modes of oscillation can be 
most clearly identified by spectral resolution of the mass moments.
A detailed frequency spectrum of the principal modes for 
gyroscopic excitation is shown in figure \ref{figure_7} as a function 
of interaction strength $C$. 
 
\subsection{Vortex precession}

Consider firstly the scissors $\langle yz \rangle $ or 
 $\langle xz \rangle $ mode of 
oscillation; the quadrupole moment results are presented in figure  \ref{figure_3} 
and the corresponding frequency spectrum in figure \ref{figure_4}. Data obtained 
using the direct solution of the time-dependent Gross-Pitaevskii equation 
are  compared with the coherent hydrodynamic calculation, 
and the 
classical top results. 
The parameters
$C=1000$ and $\lambda=\sqrt{7}$ with $N_0$ atoms and $\theta_0=4^{\circ}$ 
translate to : $I_s = (6/7)\mu_{\rm TF} N_0$,
$I_{xx}=(16/49) \mu_{\rm TF}N_0$ and  $I_{zz} = (4/7)\mu_{TF} N_0$. 
Considered as a classical rigid top, the condensate, viewed 
along the $x$-direction will exhibit quadrupole oscillations due to 
nutation. If the top  
precesses then this  moment vanishes each
 time the axis nutates along the $xz$-plane. 
Using equation (\ref{qclass}), 
the classical moment $\langle yz\rangle$  oscillates 
with a carrier 
frequency $\omega_p = (49/32) \mu_{\rm TF}^{-1} \approx 0.185 \omega_0$, 
and carrier  amplitude
 $\langle yz \rangle_{\rm max} = 2(I_{zz}-I_{xx}) \theta_0 \approx  0.283$.
The classical prediction for the beat period of the
 $\langle yz \rangle$ quadrupole,  is in fairly good agreement with 
the unperturbed coherent hydrodynamic results 
$\omega_p={\textstyle{1 \over 2}}(3-\sqrt{7})\omega_0 \approx 0.177 \omega_0$ 
as shown in figure \ref{figure_3}. Neither of these 
results compare well with the  more accurate lower-frequency 
result from the 
Gross-Pitaevskii equation: $\omega_p =0.137\omega_0$. The  lower
beat (precession) frequency  (figure \ref{figure_3}) 
contrasts with  higher 
mode frequencies compared with the hydrodynamic limit, 
as illustrated in figure \ref{figure_4}. In figure \ref{figure_7} 
it is clear that, for $C=1000$, the 
curves (1) and (2) corresponding to these modes have
not reached the asymptotic limit $C \rightarrow \infty$.

Very  recently measurements of  vortex precession \cite{ho02}
in a trap were made under the following conditions $\lambda \approx 2.83$,
$ \omega_0 = 2\pi \times 62 $ Hz,
 and $N_0 \approx 19000\pm 4000$ atoms. This 
corresponds to an interaction strength $ C \approx 2870 \pm 600 $ which 
is within  the hydrodynamic regime but should exhibit  quantal effects.  
The classical precession frequency 
is  $\omega_p^c = 0.120 \pm 0.012 \omega_0 $ where the error is due to the 
uncertainty in $N_0$. A calculation using the 
Gross-Pitaevskii equation gives  $\omega^{\rm  GPE}_p \approx 0.104 \pm 0.08 \omega_0$. 
As in the case previously discussed ($\lambda=\sqrt{7}$, $C=1000$)
 this  rate of 
precession is slightly lower than the classical/hydrodynamic prediction.
The equivalent frequencies for the vortex precession are 
 $f^{\rm  GPE}_p\approx 6.5 \pm 0.5 {\rm Hz}$ 
and   $ f^{c}_p\approx 7.4 \pm 0.5$Hz. Measurements 
were performed by imaging the vortex line in the horizontal plane 
and  trap tilting along both the $xz$ and $yz$ planes and 
were in good agreement. 
Taking the average of the results gives the experimental 
estimate  $f^{\rm exp} = 7.8 \pm 0.6 {\rm Hz}$; a slightly faster rate of precession 
than the quantal prediction, and more 
in line with  hydrodynamic theory. 

While the limitations of zero temperature 
models, as described in this paper, are well known, the hydrodynamic approximation 
is generally less accurate than the Gross-Pitaevskii equation. Moreover, 
important effects neglected by this model are   the viscous effects of
 the thermal fluid, as observed  \cite{ho02}. However, these effects 
 would probably reduce 
 $\omega^{\rm  GPE}_p$
further and thus increase the gap between  experiment 
and theory.
 It has been suggested  \cite{ho02} that two other effects might have a role in 
explaining the differences. 
Firstly, the possibility of off-centre precession 
\cite{bj} and secondly, the influence 
of additional edges vortices \cite{vort4} might have 
 significant effects. While the results of the observations and 
the theory outlined above are in very good agreement, there 
is a considerable margin for improving the simulations to study the 
contribution of these effects and of the influence of 
thermal damping.

Considering some of the other modes, we note that this 
scissors excitation of the vortex state 
weakly populates  a low-frequency mode at $\omega=1.201$; 
on figure \ref{figure_4} and labelled line (3), and 
also marked on figure \ref{figure_7}.
 This is  
the lowest $z$-dipole mode $(i;0,0,1)/(h;0,-1,1)$.  The variation 
of frequency of the mode is given in table \ref{table_1} and 
drawn in figure \ref{figure_7}.
Checks on the numerical accuracy of the 
numerical methods were also provided 
by studying  the centre-of-mass modes 
whose  frequencies remains constant
as the particle number or interaction strength $C$ varies; a  
consequence of the Kohn theorem \cite{kohn1,kohn2}, figure \ref{figure_7}. 
We note the labelling of axial excitation modes   $(i;0,2,1) \rightarrow (h;0,+1,1)$ and 
$ (i;1,0,1) \rightarrow  (h;1,-1,1)$,
 reflects the quadrupole symmetry in the hydrodynamic regime. 
 Values of the $(h;0,+1,1)$ mode frequency are given in table \ref{table_1}, 
 in which the highly accurate Bogoliubov de-Gennes results are compared with 
results using the time-dependent linear response method \cite{dermot02}. 
This confirms the accuracy both of the direct time-dependent method 
and the linear response approximation.

\subsection{Kelvin modes}

Kelvin modes are self-induced helical waves of a vortex line
and have a well-established dispersion relation for 
infinite homogeneous fluids \cite{donn}. However, in a 
trapped condensate, 
the finite boundaries means these modes are quantised, 
and the inhomogeneity of the condensate 
means that the fluid pressure  varies along the vortex line. 
The signature of the Kelvin modes is the distortion of the vortex line 
and can be detected in the gyroscope motion. 
The lowest-order and most strongly-coupled modes, $m=\pm 1$,
are excited  by the small angle rotation of the trap. Since 
the modes $m=+1$ and $m=-1$ are populated in approximately equal 
weights this   leads to a kink in the vortex   in the 
vertical plane.
The bending and twisting of the vortex line is more pronounced
at the low-density low-pressure edges of the condensate (figure \ref{figure_2}). 
Analysis of the horizontal plane quadrupole moment $\langle xy \rangle $ shows 
periodicity  but  irregularity  (figure \ref{figure_5}). 
The two-mode classical theory, equation (\ref{qclass}), strongly 
disagrees with the quantal results shown in  figure (\ref{figure_5}). 
The amplitude 
of oscillation is much larger than classically predicted, and is
associated with the vortex $s$-shape outward bend. However  the main 
feature is the  quantal oscillations
that indicate twisting, that is  bending of the vortex  
projected  onto the $\langle xy \rangle $ plane. 
The horizontal-plane spectrum (figure \ref{figure_6}) 
is dominated by two pairs of 
lines labelled (a,b) and (c,d). 
Figure \ref{figure_7} and table \ref{table_1} indicate the
frequency variation, with atom number $C$, of the helical modes. 
The lines (c) and (d) are the quadrupole 
pair $(h;1,+2,0)$ and $(h:1,-2,0)$, respectively. 
In the hydrodynamic limit with $\kappa=0$ 
they corresponds to the doubly-degenerate $xy$-quadrupole with 
frequency, $\omega=\sqrt{2}$.

The correlation of the pair (c) and (d)  to the ideal gas states $(i;1,3,0)$ 
and $(i;1,-1,0)$,
is shown in figure \ref{figure_6}. 
The splitting of this quadrupole 
pair $\Delta\omega= 0.443 \omega_0 $  is a signature of 
vortex precession, though 
again it is substantially smaller than the classical 
prediction $4 \omega_p \approx 0.548 $ associated with 
a rigid vortex crossing 
the $x=0$ and $y=0$ planes during  precession.
The dominance of mode (c) over (d) is apparent 
(figure \ref{figure_6}) and means the vortex 
has a positive helicity of $m=+2$.
Indeed, the strong coupling of $m=2$ states with vortex motion has been 
used to measure 
the angular momentum of the vortex state \cite{vort4}. 
The pair of high-frequency lines (a) and (b) also contribute strongly to the 
spectrum. These are the dipole-like modes $(h;1,1,0)$ and $(h;2,-1,0)$ 
, degenerate as $C \rightarrow 0$ with symmetry $(i;1,2,0)$ and $(i;2,0,0)$. 
The pair of lines are of approximately equal weighting and therefore  cancel
helicities. As $C\rightarrow \infty $, these  
lines converge slowly towards the common limit $\omega=2.321$. 
 Finally the spectrum shows two low-intensity lines at 
$\omega \approx 0.274 $ and $\omega \approx 1.892$. These are not 
new modes but rather vestiges 
of the 
$\langle yz \rangle$ modes; the low frequency
line being the  beat frequency corresponding to the 
splitting of lines (1) and (2) 
in figure \ref{figure_4}, 
and the higher frequency line, the frequency difference 
of  modes (1) and (3).

\section{Conclusions}

In conclusion, we have studied the gyroscopic dynamics of a trapped
Bose-Einstein condensate containing a vortex.  We modelled the system using
a classical top description and numerically by solution of the
time-dependent Gross-Pitaevskii equation. We also compared our results
with the hydrodynamic approximation.
The linear-response equations for the system were solved giving the
 excitation
spectra and  amplitudes. The superfluid precession and nutation of 
the vortex were
found to display quantum frequency shifts. 
The precession 
frequency was calculated and found to be 
consistent with recent experiments, though a small 
discrepancy exists. Vortex bending and 
twisting modes of excitation were observed.  We found, in the scissors 
excitation of the vortex state, that vortex bending 
in the vertical plane is associated with the $m=\pm 1$ 
modes of oscillation, while the 
helical oscillation, vortex twisting, is 
dominated by the  $m=+2$  Kelvin mode.

We are very grateful to Profs. C. J. Foot (Oxford) and J.-P. Hansen (Bergen) 
for many helpful discussions and 
exchanges on this problem.
We gratefully 
acknowledge the support of the Bergen Computational Physics
Laboratory in the framework of the European Community -
Access to Research Infrastructure action of 
the Improving Human Potential Programme; 
Dermot McPeake thanks the Department for Employment and Learning
Northern Ireland for financial support through the provision of
a Postgraduate Studentship.
Halvor M{\o}ll Nilsen acknowledges support from the Norwegian
Research Council.

\bibliographystyle{plain}

\end{document}